# Three-dimensional visible-light invisibility cloak

Bin Zheng[1,2], Rongrong Zhu[1,2], Liqiao Jing[1,2], Yihao Yang[1,2], Lian Shen[1,2,*], Huaping Wang[1,3], Zuojia Wang[4], Xianmin Zhang[2], Xu Liu[1], Erping Li[2] and Hongsheng Chen[1,2,*]

[1] State Key Laboratory of Modern Optical Instrumentation, The Electromagnetics Academy at Zhejiang University, Zhejiang University, Hangzhou 310027, China

[2] Key Laboratory of Micro-Nano Electronics and Smart System of Zhejiang Province，College of Information Science & Electronic Engineering, Zhejiang University, Hangzhou 310027, China

[3] Institute of Marine Electronics Engineering, Zhejiang University, Hangzhou, 310058, China

[4] School of Information Science and Engineering, Shandong University, Jinan 250100, China

Corresponding author: L. S. (lianshen@zju.edu.cn), H. C. (hansomchen@zju.edu.cn)

**Abstract**

The concept of an invisibility cloak is a fixture of science fiction, fantasy, and the collective imagination. However, a real device that could hide an object from sight in visible light from absolutely any viewpoint would be extremely challenging to build. The main obstacle to creating such a cloak is the coupling of the electromagnetic components of light, which would necessitate the use of complex materials with specific permittivity and permeability tensors. Previous cloaking solutions have involved circumventing this obstacle by functioning either in static (or quasi-static) fields where these electromagnetic components are uncoupled or in diffusive light scattering media where complex materials are not required. In this paper, we report concealing a large-scale spherical object from human sight from three orthogonal directions. We achieve this result by developing a three-dimensional (3D) homogeneous polyhedral transformation and a spatially invariant refractive index discretization that considerably reduce the coupling of the electromagnetic components of visible light. This approach allows for a major simplification in the design of 3D invisibility cloaks, which can now be created at a large scale using homogeneous and isotropic materials.

**Introduction**

Complex creatures, including humans, rely on their senses to obtain information about and react to their living environments. Visual illusions can affect the behavior of such creatures by deceiving their subjective judgment. Invisibility, one of the ultimate visual illusions, was almost inconceivable prior to the ingenious theory of transformation optics, which was first proposed in 2006.[1,2] Using transformation optics, a cloak could theoretically render an object invisible by guiding light around it as if nothing was there.

Despite its elegance, the proposed general cloaking theory[1-8] is tremendously difficult to implement; as a result, all previous implementations have involved special cases.[9-18] The first experimental validation of an invisibility cloak was a two-dimensional (2D) microwave cloak[9] for single-frequency transverse electric (TE) waves. A non-magnetic metamaterial cloak for transverse magnetic (TM) waves at microwave frequencies was subsequently proposed.[10] Then, carpet cloaks were introduced; such cloaks are easier to implement because they are based on the principle of hiding an object under a reflective ground plane in a semi-space.[5] Carpet cloaks have been experimentally demonstrated in the microwave range[19-23] and in the optical spectrum.[24-28] However, carpet cloaks operate by manipulating reflected light to map an object to a "flat" surface instead of making an object "disappear" from transmitted light. To date, most experiments to verify that objects can be made to disappear from plain sight have been restricted to 2D cloaks because three-dimensional (3D) cloaks are particularly challenging.

Barriers to the implementation of 3D cloaks are associated with the complex anisotropic parameters required to simultaneously achieve both permittivity and permeability. In the case of a 2D cloak, such as the cylindrical cloak shown in **Figure 1a**, waves propagate in a plane in which the TE and TM polarizations can be decoupled, which reduces complexity. For a TE wave with an electric field only in the z-direction, this cloak requires only certain components of the permittivity and permeability tensors ($\mu_\rho$, $\mu_\theta$ and $\varepsilon_z$); conversely, for a TM wave with a magnetic field only in the z-direction, the required components of these tensors are $\varepsilon_\rho$, $\varepsilon_\theta$ and $\mu_z$. However, for a 3D cloak, such as the spherical cloak shown in Figure 1b, waves propagate in 3D space, where their polarizations are coupled and are not easy to define. Thus, a cloak must satisfy requirements for all parameters and is difficult to achieve. In certain special cases, such as situations involving d.c. and quasi-static spectra, the electric and magnetic fields can still be decoupled, and a 3D cloak can be attained by considering only permeability.[29] In the regime of light diffusion, which leads to multiple light scattering, a 3D cloak can be achieved via the addition of diffusive shells.[18] However, until now, a 3D cloak that functions for plain sight at optical frequencies had remained an unsolved problem.

**Results and Discussion**

Here, we achieve a 3D cloak for visible light by performing a 3D polyhedral transformation and using an approach involving a spatially invariant refractive index discretization. Our experimental demonstration shows that this cloak, which is made of isotropic materials, can hide macroscopic objects in fully polarized visible light. In contrast with a previous

2D cloak for the visible-light spectrum ,[16] our cloak is effective for different viewing angles in 3D space.

For a spherical cloak, light travels in an arc inside the cloak (**Figure 2a**), and spatially inhomogeneous constitutive parameters are needed. Unlike traditional approaches, our polyhedral transformation method can be used to avoid this inhomogeneity. In fact, the use of different segments constructed with homogeneous constitutive parameters allows the trajectory of light to be controlled in a different way. Incident light bends at the interfaces of different segments in the cloak but can still perfectly bypass the hidden region (Figure 2b).

Although the polyhedral transformation eliminates the inhomogeneity of the material, anisotropic permittivity and permeability are still simultaneously required; satisfying this requirement is a major challenge with current metamaterial designs and high-resolution 3D fabrication methods. Here, a spatially invariant refractive index discretization approach is used to replace the anisotropic indices with isotropic indices (see the supporting information). This approach is similar to the quasi-conformal mapping method that was used to remove anisotropic constitutive parameters for the aforementioned carpet cloak design.[5] One difference between these approaches is that the quasi-conformal mapping in the carpet cloak design results in an inhomogeneous grid[5] but the polyhedral transformation applied here generates a homogeneous grid for each segment. Moreover, unlike the carpet cloak, for which the reflection of light is considered and a line is transformed into another line, yielding moderate anisotropic parameters,[5] the spherical cloak affects transmitted light and transforms a point into an inner sphere, resulting in highly anisotropic material parameters.[1] Anisotropy is much greater for the grids in the spherical cloak than for the grids in the carpet cloak and cannot be directly ignored. Consequently, for the spherical cloak, replacing grids of highly anisotropic materials with grids of isotropic materials is no longer a valid option for omnidirectional incident light. However, our cloak can still guide light around a hidden object and make it return to its original path when the light is incident from certain 3D perpendicular directions. A detailed analysis is included in the supporting information.

As a demonstration, a cloak with a cubic shape (**Figure 3**) is fabricated. The cloak is constructed using two different types of optical glass. In particular, twelve tetrahedral pieces of glass (ZLaF78) with a permittivity of 3.61 are placed around the hidden region in the shape of a hexapod caltrop. The entire device is enclosed by eight heptahedral pieces of glass (ZBaF1) with a permittivity of 2.63, which matches the background permittivity. The regions between these glass segments are filled with water. The side length of the entire device is $L = 100$ mm, and the object to be hidden, a steel ball with a diameter of $D = 38$ mm, is placed in the device. The compression ratio is $\kappa = 0.172$ with $D_1 = 70.7$ mm and $D_2 = 27.6$ mm, respectively (see supporting information). This cloak is operational for incoherent light illumination. Additional details regarding the cloak design and the calculated trajectory of the light propagating through the cloak are provided in the supporting information.

In the experimental setup (**Figure 4a**), the cubic cloak is vertically illuminated by a laser beam with a wavelength of 532 nm. The transmission pattern of an apple is placed between the laser beam and the device to serve as scenery to verify the cloaking effect. The light that is transmitted through the cubic cloak is projected onto a black screen, and the light pattern on the black screen is captured by a digital camera behind the screen.

Figure 4b shows the performance of this cubic cloak. When the steel ball is covered by the cubic cloak, the image pattern is recovered, and the steel ball is invisible to the observer. The cases with only the background (Figure 4c) and only the steel ball (Figure 4e) have also been measured for reference; the corresponding results are shown in Figure 4d and 4f, respectively. In the former case, the laser beam is transmitted through the area of interest and directly onto the screen, and the captured image pattern is well maintained. When the steel ball is placed between the mask and the screen without the cloaking device, most of the light is blocked, the image pattern is no longer maintained, and the steel ball is visible. In contrast, the designed cubic cloak clearly performs well. The recovered patterns are slightly distorted due to deviations in fabrication and edge effects but can be improved with the application of more accurate fabrication technology.

In addition, **Figure 5** shows the performance of this cubic cloak working at the wavelength of 650 nm. The transmission pattern is a word "ZJU" and the hidden object is in a shape of hexapod caltrop (Figure 5a). The captured images are shown in Figure 5b to 5d, respectively, which indicate the broadband cloaking performance for the designed cubic cloak.

It should be noted that in our demonstration, instead of using anisotropic materials, which are always highly dispersive, we use isotropic materials with very little dispersion; such materials can therefore be effective for the entire visible spectrum. Furthermore, since the cubic cloak is symmetric in three orthogonal directions in 3D space, this device works equally well in the three directions normal to the surfaces of the cubic cloak.

**Conclusions**

In conclusion, we propose a 3D cloaking device that can hide a macroscopic object from plain sight. Compared with 2D devices, our device offers an additional degree of freedom with respect to observation angles. Our work provides a new solution for hiding an object in 3D natural illumination for the entire spectrum of human eye sensitivity and will have practical applications in surveillance technology and for security- and defense-related purposes.

**Acknowledgements**
We thank P. Rebusco for the critical reading and editing of our manuscript. This work was sponsored by the National Natural Science Foundation of China under grant nos. 61625502, 61574127, 61601408, 61775193 and 11704332; the ZJNSF under grant no. LY17F010008; the Top-Notch Young Talents Program of China; the Fundamental Research Funds for the Central Universities; and the Innovation Joint Research Center for Cyber-Physical-Society System.

**Figures**

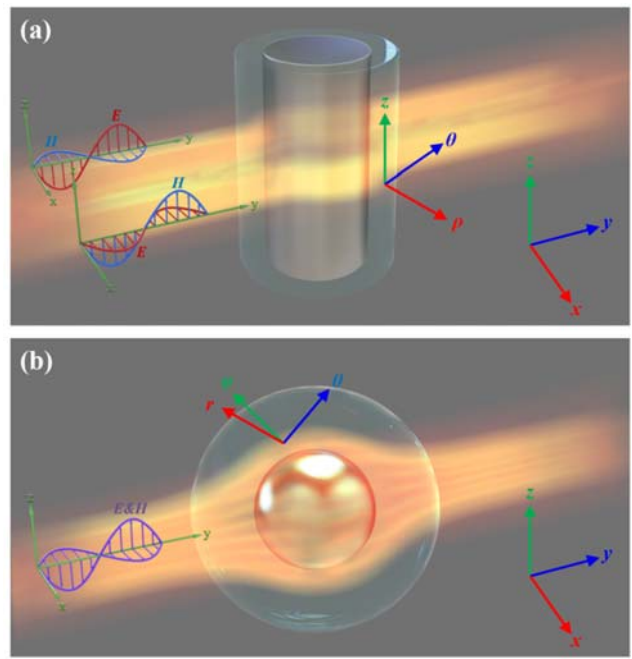

**Figure 1.** (a) 2D cylindrical cloak for which waves with TE and TM polarizations are decoupled. (b) 3D spherical cloak for which the polarizations of waves are coupled and are not easy to define.

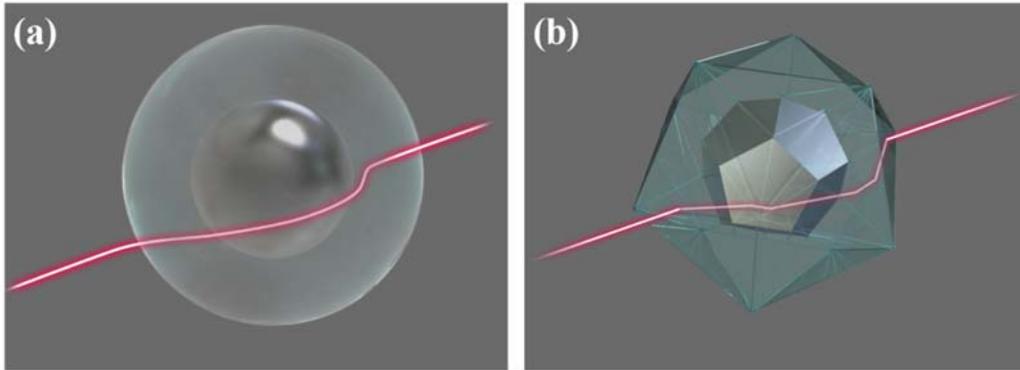

**Figure 2.** (a) Spherical cloak that guides light smoothly around the hidden region. (b) Polyhedral cloak that bends light at the boundaries of different segments to perfectly bypass the hidden region.

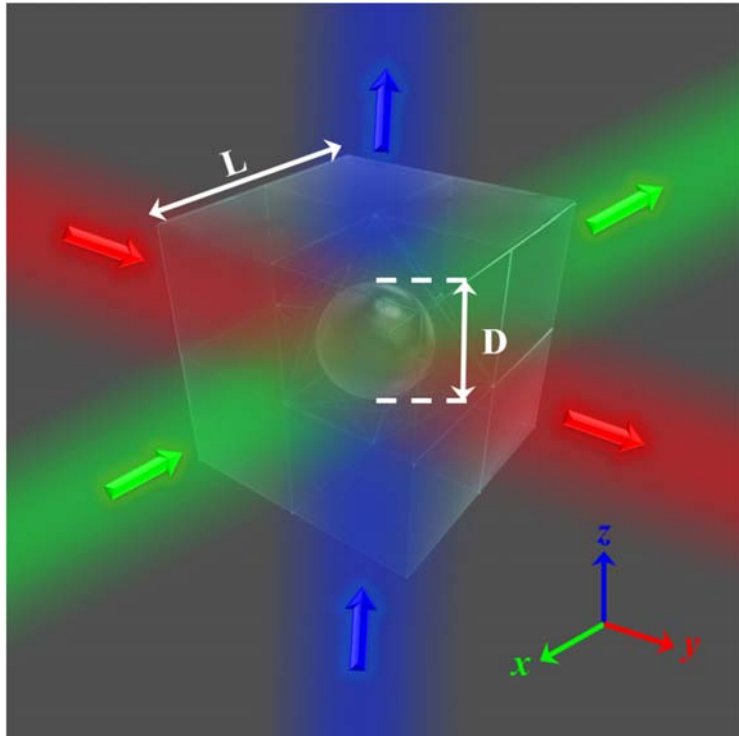

**Figure 3.** Schematic diagram of the simplified cubic cloak. This cloak is effective in three orthogonal directions in 3D space.

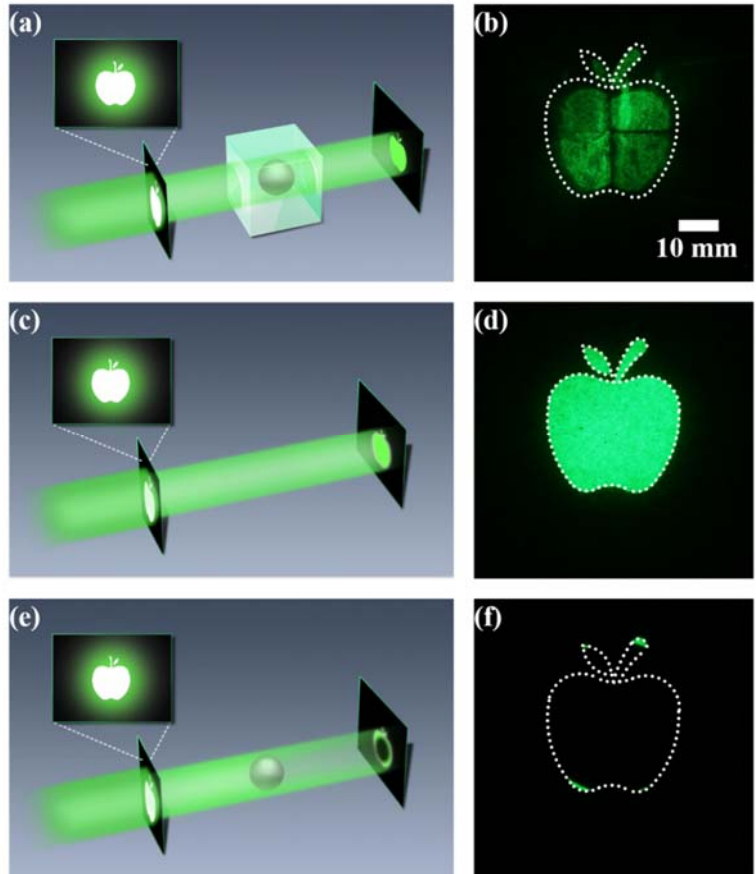

**Figure 4.** (a) Experimental setup for measuring the cloaking effect. (b) Captured image pattern for light passing through the steel ball covered with the designed cloaking device. The dotted line represents the outline of the mask's image pattern. (c) Reference case with only the background and (d) the corresponding captured image pattern. (e) Reference case with only the steel ball and (f) the corresponding captured image pattern.

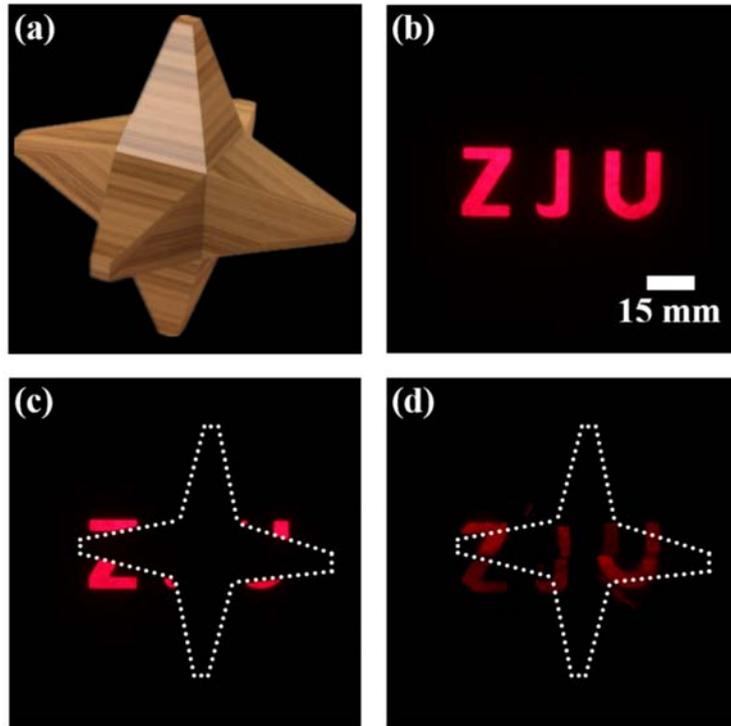

**Figure 5.** The cloaking effect of the designed cloaking device at different frequency. (a) The object to be hidden with a shape of hexapod caltrop. (b) The captured image of the reference case with light transmit directly from a transmission pattern of word "ZJU". (c) The captured image of the reference case with light transmit through the object. (d) The captured image pattern for light passing through the object covered with the designed cloaking device. The dotted line represents the outline of the hidden object.

# Supporting Information for
# Three-dimensional visible-light invisibility cloak

*Bin Zheng, Rongrong Zhu, Liqiao Jing, Yihao Yang, Lian Shen,\* Huaping Wang, Zuojia Wang, Xianmin Zhang, Xu Liu, Erping Li and Hongsheng Chen\**

# 1. Polyhedral transformation for an ideal octahedral invisibility cloak

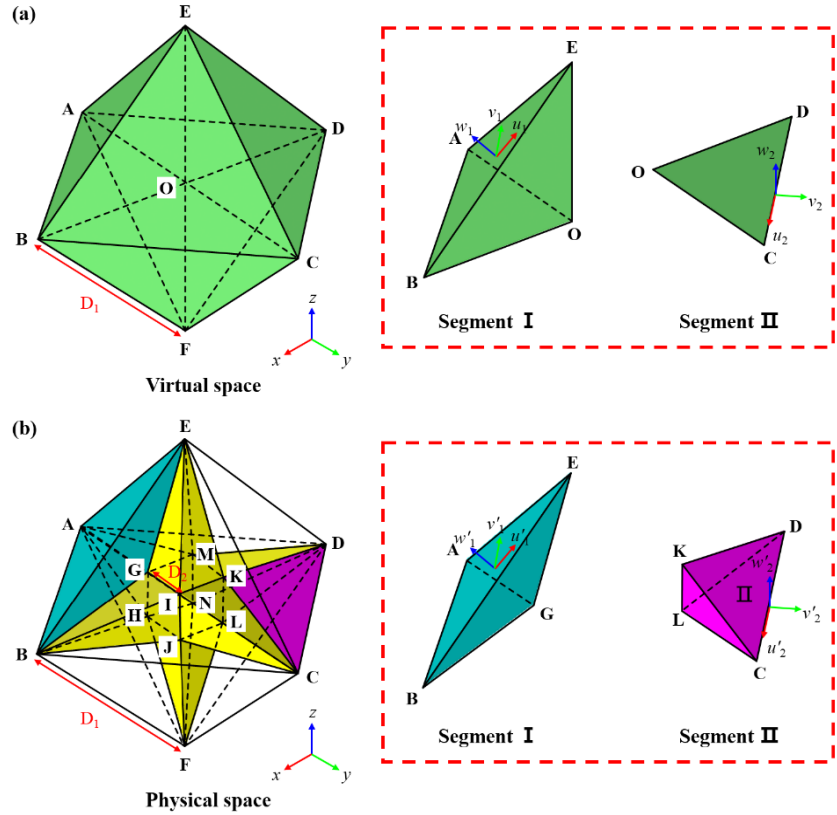

**Figure S1.** Schematic of a polyhedral transformation to design the cloak with (a) virtual space and (b) physical space. Each segment is transformed along its local coordinate axes. The segment of region Ⅰ is transformed from a tetrahedron to another tetrahedron, while the segment of region Ⅱ is transformed from a triangle to a tetrahedron. The hidden region is a hexapod caltrop shaped region which consists of one hexahedron and six pyramids. The side length of the outer octahedron is $D_1$ and the side length of the inner hexahedron is $D_2$.

In order to better illustrate our octahedral cloak design, we begin with a virtual space in the shape of an octahedron in free space, as shown in **Figure S1**a. The space is divided into 8 tetrahedral segments OABE, OBCE, OCDE, ODAE, OABF, OBCF, OCDF and ODAF (region I). There are also 12 triangular interfaces between every two adjacent tetrahedral segments, namely OAB, OBC, OCD, ODA, OAE, OBE, OCE, ODE, OAF, OBF, OCF and ODF (region II). The transformation is applied to each segment along its corresponding local coordinate axes, with compression or extension, leading to the physical space shown in Figure S1b. The segments of region I are transformed into the tetrahedral segments GABE, IBCE, KCDE, MDAE, HABF, JBCF, LCDF and NDAF, while the segments of region II are transformed into the tetrahedral segments ABGH, BCIJ, CDKL, DAMN, AEGM, BEGI,

CEIK, DEKM, AFHN, BFHK, CFJL and DFLN, respectively. The inset illustrates the transformation in detail, by showing the specific segments from each region. After the transformation, a hidden region in the shape of a hexapod caltrop is generated (yellow region in Figure S1b).

The transformation functions applied for different segments are:

$$u'_1 = u_1, v'_1 = v_1, w'_1 = \kappa w_1, \text{ for segments in region I}$$
$$u'_2 = u_2, v'_2 = \kappa_a v_2, w'_2 = \kappa_b w_2, \text{ for segments in region II} \quad (S1)$$

where, $\kappa$, $\kappa_a$ and $\kappa_b$ are the compression or extension ratios with $\kappa = \left(\frac{\sqrt{6}}{6}D_1 - \frac{\sqrt{3}}{2}D_2\right) / \left(\frac{\sqrt{6}}{6}D_1\right)$, $\kappa_a = \left(\frac{1}{2}D_1 - \frac{\sqrt{2}}{2}D_2\right) / \left(\frac{1}{2}D_1\right)$ and $\kappa_b = \infty$, respectively.

The constitutive parameters are:

$$\varepsilon'_{1u} = \mu'_{1u} = \frac{1}{\kappa}, \varepsilon'_{1v} = \mu'_{1v} = \frac{1}{\kappa}, \varepsilon'_{1w} = \mu'_{1w} = \kappa, \text{ for segments in region I}$$
$$\varepsilon'_{2u} = \mu'_{2u} = \frac{1}{\kappa_a \kappa_b}, \varepsilon'_{2v} = \mu'_{2v} = \frac{\kappa_a}{\kappa_b}, \varepsilon'_{2w} = \mu'_{2w} = \frac{\kappa_b}{\kappa_a}, \text{ for segments in region II} \quad (S2)$$

## 2. Spatial invariant refractive indices discretization

In Equation (S2), the parameters for segments in regions I and II are all diagonal tensors in their local coordinate systems. We denote $g_u$, $g_v$ and $g_w$ the principal values of the permittivity and permeability tensors, and we define the corresponding refractive indices to be $n_u = \sqrt{g_v g_w}$, $n_v = \sqrt{g_u g_w}$ and $n_w = \sqrt{g_u g_v}$. The refractive indices for each segment are:

$$n'_{1u} = 1, n'_{1v} = 1, n'_{1w} = \frac{1}{\kappa}, \text{ for segments in region I}$$
$$n'_{2u} = 1, n'_{2v} = \frac{1}{\kappa_a}, n'_{1w} = \frac{1}{\kappa_b}, \text{for segments in region II} \quad (S3)$$

The anisotropy factor --- i.e., the maximum ratio of the refractive indices in different directions --- for each segment are calculated to be $\alpha_1 = \frac{1}{\kappa}$ (for segments in region I) and $\alpha_2 = \frac{\kappa_b}{\kappa_a}$ (for segments in region II), respectively. Since the grid obtained from the polyhedral transformation is homogeneous, the anisotropy factor in each segment is also homogeneous and it does not vary in space.

The anisotropy in the design of the octahedral cloak is eliminated with a spatial invariant refractive indices discretization approach. This approach is similar to the quasi-conformal mapping method that has been used in the carpet cloak design[1] to minimize the average anisotropy factor in space. In that approach, the anisotropic refractive indices were replaced by isotropic ones, and the carpet cloak could be fabricated using dielectrics with an inhomogeneous refractive indices distribution. Although this approach leads to some lateral shifts,[2] the cloak could achieve an omnidirectional cloaking effect in a broad frequency band regardless of the light polarization.

In the designed octahedral cloak, since the anisotropy factors in the segments are not small enough, this approach influences the cloaking effect. Nevertheless, it is still valid for certain incident angles. We therefore replace the segments in the octahedral cloak with isotropic materials, as shown in **Figure S2**. The segments discretized with these spatial invariant refractive indices guide the light as designed in certain directions. For example, assume the incident light is propagating along the –z direction with $(\theta_0, \varphi_0) = (\pi, 0)$; then the isotropic segments with refractive indices

$$n_0 : n_I : n_{II} = 1 : \sqrt{\frac{2+\kappa^2}{3}} : 1+\kappa \tag{S4}$$

keep the same trajectories as the anisotropic ones.

However, from Equation (S4), we found that $n_I$ should be smaller than unit, which is hard to realize in practice. This issue can be solved by introducing an additional type of segments (segments III), in the shape of heptahedrons, to act as background. As shown in Figure S2b, the three squared faces of each segment III are perpendicular to the x-axis, y-axis and z-axis, respectively. In this way, the ray trace of the light will still remain unchanged and the whole cloaking device can be effective in plain sight. With this addition, the refractive indices of segments I, II and III are:

$$n_I : n_{II} : n_{III} = 1 : \sqrt{\frac{3(1+\kappa)^2}{2+\kappa^2}} : \sqrt{\frac{3}{2+\kappa^2}} \tag{S5}$$

In the practical realization, the side length of the entire device is $L = 100$ mm, with $D_1 = 70.7$ mm and $D_2 = 27.6$ mm, respectively. The compression ratio $\kappa$ is then calculated to be $\kappa = 0.172$ and the refractive indices $n_I : n_{II} : n_{III} = 1.333 : 1.9 : 1.621$. In addition, the volume of the total hidden region (a hexapod caltrop shaped region which consists of one hexahedron and six pyramids) is $V_{total} = \sqrt{2} D_1 D_2^2$, which is about 7.6% compare with the volume of the entire device.

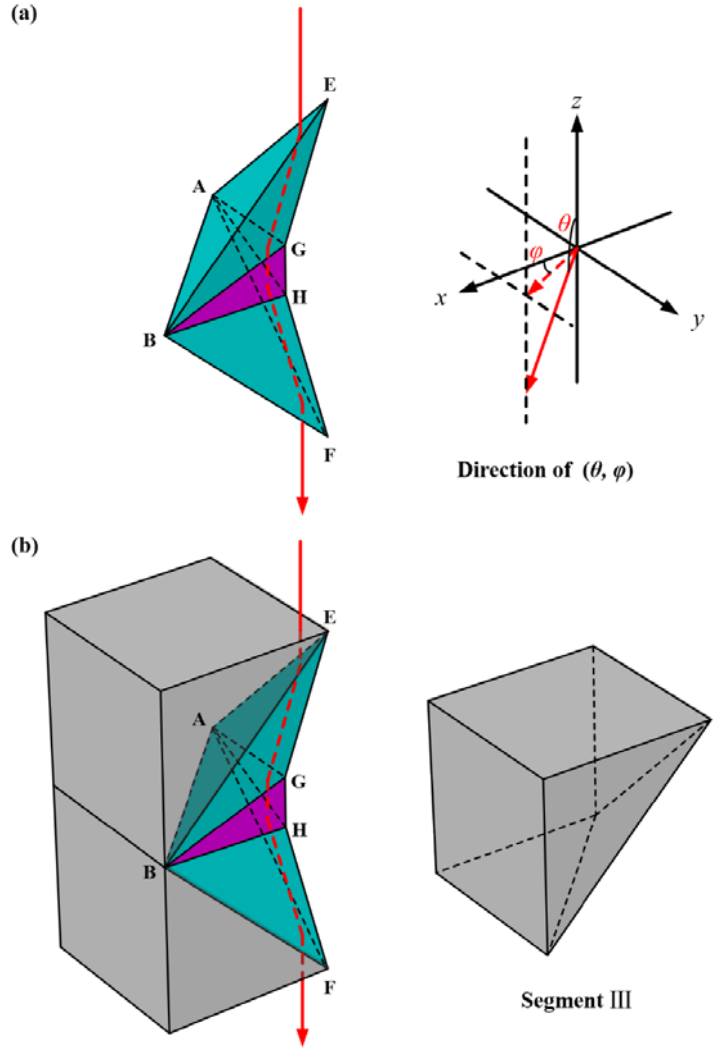

**Figure S2.** (a) Light propagating along the –z direction through specific segments. Each segment can be replaced by an isotropic material leaving the trajectory of light unchanged. (b) An additional type of segments (segments III) is applied to act as background. These segments do not affect the trajectory of light for light propagating along the main axes directions.

It should be noted that Figure S2a shows only part of the device, but due to its high symmetry, the approach will still be valid for light propagating along the –z direction from other parts of the device. In addition, for different directions $(\theta_0,\varphi_0)=(\pi/2,\pi)$ and $(\theta_0,\varphi_0)=(\pi/2,-\pi/2)$ --- i.e., along the –x and –y directions, respectively --- the device works equally well with the same refractive indices. Hence, the cloak simplified with this discretization approach can be effective for light that propagates along the main axes directions, where it is the most important in 3D space.

## 3.   Calculation of the trajectory of light

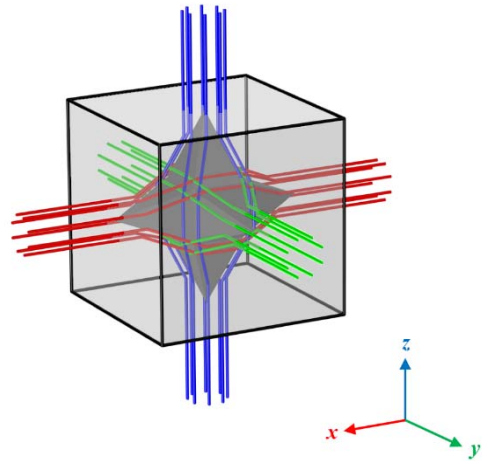

**Figure S3.** Calculated light ray trajectories in the simplified cloaking device.

The calculated trajectories of light in the simplified cloaking device are shown in **Figure S3**, where the red, green and blue lines represent light incident from different axes. For simplification, here the inner details of the cloaking device are omitted and only the outline of the simplified cloak is shown. Not only the light rays incident along the x-axis bypass the hidden object, but furthermore the trajectories of light in the simplified cloak are strictly the same as in the ideal one. Due to the structure symmetry, this simplified cloak is also effective for light incident along the y-axis and z-axis.

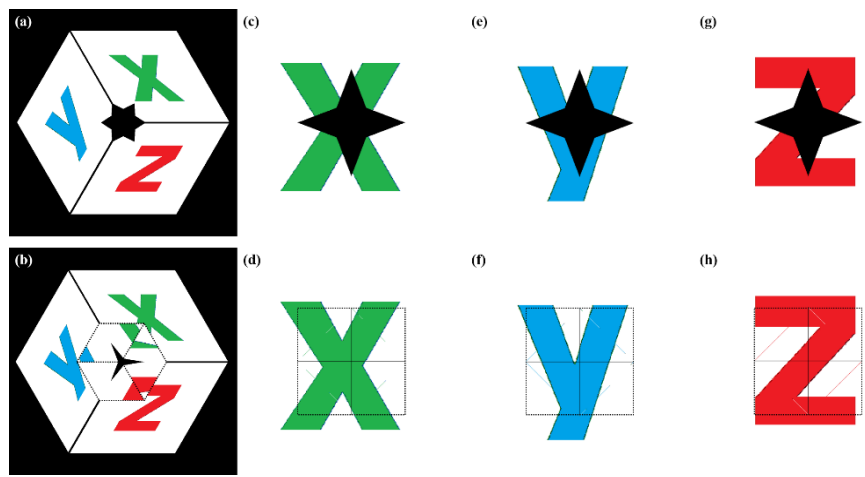

**Figure S4.** Virtual images obtained with numerical simulation. (a) The defined background image with an object placed in the middle. (b) The object is enclosed by the cloaking device. (c-h) Observing angles along the direction of the three orthogonal axes, without and with the cloaking device (top and bottom respectively). The dotted lines represent the outline of the cloaking device.

We performed a numerical simulation to validate the effectiveness of our cloaking device. As shown in **Figure S4**a, we define the background images along the direction of the x-axis, y-axis and z-axis, with the hidden object placed in the middle. All the segments of the cloak have been defined with shapes and refractive indices as discussed earlier. As shown in Figure S4b, the algorithm will display a virtual image by calculating the trajectory of light along a specific viewing angle. This algorithm does not take into account reflection, therefore the light blocked or totally reflected by the object will be shown in black.

Figures S4c to S4h show the virtual images calculated with this algorithm for viewing angles along the direction of the three orthogonal axes, without and with the cloaking device. We can see from these simulations that the background images are blocked by the object, but they are recovered when the cloaking device is in use, proving that our proposed cloaking device can be effective, for these specific angles, to guide the blocked light rays around the hidden object.